**Ligand-Hole in SnI$_6$ Unit and Origin of Band Gap in Photovoltaic Perovskite Variant Cs$_2$SnI$_6$**

Zewen Xiao,[*,1,2] Hechang Lei,[2,3] Xiao Zhang,[2] Yuanyuan Zhou,[4] Hideo Hosono,[1,2]

and Toshio Kamiya[*,1,2]

[1]Materials and Structures Laboratory, Tokyo Institute of Technology, Yokohama 226-8503

[2]Materials Research Center for Element Strategy, Tokyo Institute of Technology, Yokohama 226-8503

[3]Department of Physics, Renmin University of China, Beijing 100872, China

[4]School of Engineering, Brown University, Providence, RI 02912, United States

Cs$_2$SnI$_6$, a variant of perovskite CsSnI$_3$, is gaining interest as a photovoltaic material. Based on a simple ionic model, it is expected that Cs$_2$SnI$_6$ is composed of Cs$^+$, I$^-$, and Sn$^{4+}$ ions and that the band gap is primarily made of occupied I$^-$ 5$p^6$ valence band maximum (VBM) and unoccupied Sn$^{4+}$ 5$s$ conduction band minimum (CBM) similar to SnO$_2$. In this work, we performed density functional theory (DFT) calculations and revealed that the real oxidation state of the Sn ion in Cs$_2$SnI$_6$ is +2 similar to CsSnI$_3$. The +2 oxidation state of Sn originates from 2 ligand holes ($\underline{L}^+$) in the [SnI$_6$]$^{2-}$ octahedron unit, where the ligand [I$_6$] cluster has the apparent [I$_6^{6-}\underline{L}^+_2$]$^{4-}$ oxidation state, because the band gap is formed mainly by occupied I 5$p$ VBM and unoccupied I 5$p$ CBM. The +2 oxidation state of Sn and the band gap are originated from the intracluster hybridization and stabilized by the strong covalency of the Sn–I bonds.

Lead halide perovskites invoked new development research in the third-generation photovoltaics because of the high power conversion efficiencies (PCEs) of those photovoltaic cells as high as 20.1%.[1–6] These compounds are represented by the general chemical formula $ABX_3$ (A = Cs, $CH_3NH_3$, or $CH_2NH=CH$; B = Pb or Sn; X = I, Br or Cl), where the A cations are located in the cubic network of corner-sharing $BX_6$ octahedra, as shown in Figure 1a. However, these compounds and their devices suffer from the long-term instability in an ambient atmosphere, and the water-soluble toxic lead components are potential risk for the environment issue. Therefore, Sn derivatives of perovskite compounds, $ASnX_3$, have been examined.[7–9] However, also these $ASnX_3$ perovskites are very sensitive to the ambient atmosphere (oxygen, moisture, etc.).[7–11] Then, a class of perovskite variant $A_2SnX_6$[12] is recently suggested by Lee et al.[13] as air-stable alternatives in the application to solar cells. $Cs_2SnI_6$ is a typical example and has shown better stability in the ambient environment and reasonably high PECs up to 7.8% in $Cs_2SnI_6$-based solar cells.[13] Based on a simple ionic model, Lee et al.[13] claimed the improved air-stability of $Cs_2SnI_6$ is owing to the +4 oxidation state in this compound. However, the determination of oxidation states is not simple and unique in particular for metal cations that take multiple stable oxidation states; e.g., there have been reports on unusual oxidation states in some complex compounds such as $CaCu_3Fe_4O_{12}$.[14] Similarly, it is not clear whether such a simple ionic $Cs^+_2Sn^{4+}I^-_6$ model is applicable for such a complex molecular iodosalt compound in which strong covalent bonds may exist. First-principles calculations have widely been employed to understand electronic structures and assign the oxidation states. However, there is ambiguity to evaluate the oxidation states of ions because there is no unique accepted method to attribute the electron density in the inter-ion spaces to each ion. Therefore, many charge analysis methods have been proposed and examined, which include Mulliken population analysis, natural bonding orbital analysis, Born effective charges, and Bader charge analysis.[15] Among them, the Mulliken population analysis has ambiguity how to assign the non-diagonal density matrix components to the bonding ions and depends significantly on the choice of the atomic orbital basis set, whose issue is more serious for plane wave (PW) basis sets.

Some PW method codes like Vienna Ab initio Simulation Program (VASP) counts the electron density within a sphere around an atom core with an artificially-given Wigner−Seitz radius ($R_{WS}$) to calculate the total electron number ($n_e$) assigned to this atom; however, the result depends largely on the choice of $R_{WS}$. Born effective charges appear to provide good values for materials with small dielectric permittivity, but gives extraordinary large values with high anisotropy for materials with large dielectric permittivity because it counts electron redistribution induced by motion of ions. Bader charge analysis divides all the space in unit cell to the constituent ions based on the Bader partitioning scheme and provides unambiguous values independent of the basis set.[15]

In this work, we investigated the electronic structure and oxidation states of $Cs_2SnI_6$ by density theory functional (DFT) calculations. Based on the simple ionic model, the electronic structure of $Sn^{2+}$-based $Cs^+Sn^{2+}I^-_3$ and $Sn^{4+}$-based $Cs^+_2Sn^{4+}I^-_6$ can be illustrated schematically by Figures 2a and 2b, respectively. While the Sn 5$s$ orbitals of $Cs^+Sn^{2+}I^-_3$ are fully-occupied,[10] those of the expected $Cs^+_2Sn^{4+}I^-_6$ model are unoccupied and contribute to the conduction band minimum (CBM), similar to the case of $SnO_2$.[16] The band gaps of both $Cs^+Sn^{2+}I^-_3$ and $Cs^+_2Sn^{4+}I^-_6$ are of a charge-transfer type band gap where the valence band maximum (VBM) and CBM are formed mainly by anions and cations, respectively. Here, we found that while the VBM consists of I 5$p$–I 5$p$ antibonding states as expected, the CBM, unexpectedly, consists of I 5$p$–Sn 5$s$ antibonding states as illustrated in Figure 2c. The Sn 5s orbitals indeed form electronic states ~7 eV deeper than the VBM and are fully occupied, indicating that Sn in $Cs_2SnI_6$ should be in the +2 oxidation state. To discuss the oxidation states of the constituent ions, we examined the above different charge analysis methods and finally chose the Bader method by comparing with related compounds with well-accepted $Sn^{2+}$ and $Sn^{4+}$ oxidation states. The calculated oxidation state of I is a bit smaller than −1 and to be −2/3, because the [$I_6$] ligand cluster is regarded to have 2 ligand holes and its oxidation state is $[I_6]^{4-}$ due to the I 5$p$–Cs 6$s$ antibonding CBM state. We will discuss the origin of the chemical stability of the $Cs_2SnI_6$ phase based on these results.

**Computational Details**

DFT calculations were performed using the projector-augmented plane wave (PAW) method as implemented in the VASP code.[17] Cs ($5s$)($5p$)($6s$), Sn ($5s$)($5p$), and I ($5s$)($5p$) are treated as valence states in the PAW potentials. The plane wave cutoff energy was set to 275.4 eV. For the exchange-correlation functional, we confirmed that the local density approximation (LDA) and the Perdew−Burke−Ernzerhof (PBE96)[18] generalized gradient approximation (GGA) functionals underestimated the band gap for $Cs_2SnI_6$ (0.13 and 0.25 eV, respectively as listed in Table 1. The experimental value is 1.26 eV[13]). On the other hand, the Heyd−Scuseria−Ernzerhot (HSE06) hybrid functional,[19] which incorporates α = 25 % of Hartree-Fock exact exchange contribution and 1−α = 75 % of PBE96 contribution, provided a better band gap value of 0.93 eV but still smaller than the experimental value. In order to discuss the chemical bonding nature and the origin of the band gap more realistically, we adjusted the α to be 34 % so as to reproduce the reported band gap value, and the HSE06 with this α value will be used for the following discussion. A primitive cell containing one formula unit (f.u.) of $Cs_2SnI_6$ and a Γ-centered 4 × 4 × 4 $k$-mesh were employed for the periodic calculations. Prior to the electronic structure calculations, variable-cell structure relaxations were performed. To reveal the origin of the band gap in $Cs_2SnI_6$, we also performed calculations for several hypothetic structures including $[I_6]^0$ cluster, $[SnI_6]^{2-}$ cluster, and $[SnI_6]^{2-}$ sublattice models. Additionally, chemical bonding analysis was carried out for $Cs_2SnI_6$ based on the crystal orbital Hamiltonian population (COHP)[20] calculated by a tight binding−linear muffin tin orbitals−atomic sphere approximation (TB-LMTO-ASA) program,[21] where positive –COHP values indicate bonding states and vice versa. Integration of the –COHP spectrum up to the Fermi level ($E_F$) yielded –ICOHP values as measure of total overlap populations (in other words, the bond order). The Bader charge analysis was carried out by the Bader program[22] with the charge density obtained by VASP.

## Results and Discussion

$Cs_2SnI_6$ crystalizes into the face-centered-cubic (fcc) $K_2PtCl_6$-type with the space group $Fm-3m$ (the anti-fluorite structure) and the lattice parameter $a$ of 11.65 Å.[23] As shown in Figure 1c, the unit cell is composed of four $[SnI_6]^{2-}$ octahedra at the corners and the face centers and eight $Cs^{2+}$ cations at the tetragonal interstitials. Alternatively, $Cs_2SnI_6$ can be regarded as a defective variant of the perovskite $CsSnI_3$ as seen by comparing Figures 1b and 1c. Figure 1b illustrates a 2×2×2 supercell of $CsSnI_3$ in which the $[SnI_6]^{2-}$ octahedra connect to each other by sharing their corners. The $Cs_2SnI_6$ structure is obtained by removing a half of the Sn atoms at each center of the $[SnI_6]$ octahedron at intervals (i.e. the edge centers and the body center in Figure 1b), and thus the corner-shared $[SnI_6]^{2-}$ octahedra in Figure 1b become isolated in $Cs_2SnI_6$ (Figure 1c). After the half of the Sn atoms are removed, the $[SnI_6]^{2-}$ octahedra shrink slightly, leading to the smaller Sn−I bond length (2.85 Å)[23] in $Cs_2SnI_6$ than that in $CsSnI_3$ (3.11 Å)[10] as well as the smaller intraoctahedral I−I bond length (4.04 Å) than that of interoctahedral I−I' bond lengths (4.20 Å) (Figure 1d). Table 1 summarizes the literature and calculated lattice parameters, bonding lengths, and band gaps. Compared with the LDA and PBE96, the HSE06 gave better results, as it often does for semiconductors.[24] In particular, when $\alpha = 34\%$, the HSE06 provided the lattice parameters and bonding lengths within 1.5% errors from the room-temperature experimental values and reproduced the experiment band gap value.

Figures 3a and 3b show the calculated band structure and the total and projected densities of states (DOSs) for $Cs_2SnI_6$ with the HSE06, respectively. The band structure of $Cs_2SnI_6$ exhibits a direct band gap of 1.26 eV at the Γ point, which is close to the experimental value.[14] The valence band (VB) consists of I $5p$ orbitals, and its band width is small (only 2.38 eV in the total width). Below the VB, there is another I $5p$ band localized between −2.90 and −3.62 eV, which is slightly hybridized with Sn $5p$ orbitals. The Sn $5s$ orbital forms a very deep band between −7.18 and −6.84 eV, and has little contribution to the VB. On the other hand, the conduction band (CB) extends from 1.26 to 2.56 eV and mainly consists of I $5p$ orbitals hybridized with Sn $5s$ orbitals. This breaks the above expectation in Figure 2b that the CBM would consist of Sn 5*s* orbitals based on the simple ionic model of $Cs^+_2Sn^{4+}I^-_6$. Upper CB bands starts from 4.86 eV and separate from the CB with a forbidden gap of 2.30 eV, which consist of Sn 5*p*, Cs 6*s*, 5*d* and I 6*s*, 5*d* orbitals. It can also be seen that the Cs ion has little contribution to the VB and the CB. The calculated electronic structure of $Cs_2SnI_6$ is represented schematically by Figure 2c. Similar to the case of $CsSnI_3$ (Figure 2a), the Sn 5*s* orbitals in $CsSnI_6$ are occupied, which indicates that the real oxidation state of Sn in $Cs_2SnI_6$ is +2, similar to that in $CsSnI_3$ and not the expected oxidation state of +4.

Based on the $Sn^{2+}$ and the $Cs^+$ ions in $Cs_2SnI_6$, the overall charge of the [$I_6$] cluster must be −4 and then each I atom equivalently has an apparent fractional oxidation state of −2/3 (i.e. the $5p^{5⅔}$ configuration), smaller than the conventional oxidation state of −1 (i.e. the fully occupied $5p^6$ configuration); this is consistent with the above electronic structure that 1/18 of $I_6$ 5*p* orbitals (i.e. the single conduction band) are unoccupied. In other words, this electronic structure is understood that the [$I_6$] ligand cluster has 2 ligand holes $\underline{L}^+$ and is represented as $[I^-_6\underline{L}^+_2]^{4-}$. This leads to converting the expected formal oxidation state of $Sn^{4+}$ to $Sn^{2+}$ via $Sn^{4+} \rightarrow Sn^{2+}\underline{L}^+_2$, being consistent with the calculation result.

The $Sn^{2+}$ and $I^{-2/3}$ states can also be perceived by the ionic distance analysis. Table 2 shows the literature bonding lengths and the ones estimated from the literature ionic radii.[25,26] The bonding lengths estimated from the ionic radii of $Cs^+$ (1.88 Å), $Sn^{2+}$ (1.02 Å), and $I^-$ (2.20 Å) explain well the actual ones in $CsSnI_3$,[10] but not in $Cs_2SnI_6$; i.e., compared with the Sn–I length estimated using the $Sn^{2+}$ radius (3.22 Å), the one using the $Sn^{4+}$ radius (2.89 Å) is closer to the actual value (2.85 Å), which seems inconsistent with the $Sn^{2+}$ state in $Cs_2SnI_6$. This apparent inconsistency is caused by the improper assumption of the $I^-$ state and its ionic radius, as the radius of $I^-$ is larger than the actual $I^{-2/3}$ anion in $Cs_2SnI_6$. The radius of the $I^{-2/3}$ anion is unknown, but it would reasonably be estimated to be 2.02 Å from the experimental I–I length in $Cs_2SnI_6$ (4.04 Å), which improves the estimated $Sn^{2+}$–I length to 3.08 Å. This value is still larger than the experimental value (2.85 Å), but the experimental Sn–I length would be shortened also by the isolation of the $[SnI_6]^{2-}$ octahedra as explained above.

To further support the +2 oxidation state of Sn, we performed Bader charge analysis for $Cs_2SnI_6$ as well as other related compounds for comparison. Figure 4 shows the Bader oxidation states ($Z$) of Sn calculated from the Bader charges ($N$). It should be noted that although all the $Z$ values of the Sn (+1.25 and +1.56, respectively, in typical $Sn^{2+}$-based compounds SnO and $SnF_2$; +2.40 and +2.71, respectively, in typical $Sn^{4+}$-based compounds $SnO_2$ and $SnF_4$) are underestimated, the similar values are obtained if the oxidation states are the same. The Sn in $Cs_2SnI_6$ has the $Z = +1.24$, close to those in SnO (+1.25), $SnF_2$ (+1.56), $CsSnI_3$ (+0.89), binary $SnI_2$ (+0.98), and $SnI_4$ (+1.16), while much smaller than those in $SnO_2$ (+2.40) and $SnF_4$ (+2.71). This result indicates that the Sn in SnO, $SnF_2$, $CsSnI_3$, $Cs_2SnI_6$, $SnI_2$, and $SnI_4$ have similar oxidation states. The similarity in the $Z$ values between $Cs_2SnI_6$ and $CsSnI_3$ would be reasonable because the coordination structures around an Sn ion consist of [$SnI_6$] octahedra and are similar to each other. Interestingly, the $SnI_4$ has the similar Z value to that in $SnI_2$ and other $Sn^{2+}$-based compounds. It implies that, unlike F, anionic I is much larger than F and other halide ions, forms more covalent bonds, and consequently its oxidation ability is not strong enough to oxidize Sn to the +4 oxidation state; $SnI_4$ is indeed known to be a covalent molecular compound with more covalent I−I bonds.[27]

We then analyze the character and strength of chemical bonds by COHP as shown in Figure 3a along with the total DOS. It can be seen that the Sn–I bonds are all bonding states below $E_F$ while antibonding ones above $E_F$. It should be noted that there are strong bonding states at ~ −7.4 and −3.2 eV, which originate from the strong covalent interaction between Sn $5s/5p$ and I $5p$ states. The strong covalent interaction also leads to the largest –ICOHP (2.41 eV/bond) for the Sn–I bond and would be the origin of the chemical stability of $Cs_2SnI_6$. In contrast, the small –ICOHP value (0.10 eV/bond) for the Cs–I bond reflects the weak covalent interaction between Cs and I. On the other hand, the –COHP spectra of I–I and I–I' bonds show similar patterns between −2.2 and 0 eV, where the DOS is contributed dominantly by the I $5p$ state. This pattern recalls the familiar molecular orbital (MO) diagrams of homonuclear molecular dimers and clusters. The large antibonding states at $E_F$ and the negative small –ICOHP values (−0.24 and −0.064 eV/bond for the I−I and I−I' bonds, respectively) imply that the MOs of I $5p$ are almost filled up and no effective bond strength remains between the I ions. The bonding natures are confirmed also by the valence electron density maps on the (200) and (400) planes, as shown in Figures 3b and c, respectively. It can be seen that the significant electron distribution in the Sn–I bonds clearly indicate the formation of Sn–I covalent bond. In contrast, there is little electrons between Cs–I, meaning that little covalent bond exist in the Cs−I bonds.

To understand the chemical bonding nature and the origin of the band gap in $Cs_2SnI_6$, we performed DFT calculations for some hypothetic structures. First we examined the electronic structure for an isolated $I_6$ octahedron (i.e., $I_6^0$ cluster), as shown in Figure 6a. According to the energy eigenvalues at the Γ point and the group theory, the 18 I $5p$ orbitals of the $I_6$ octahedra are split to 7 groups.[28] The 6 radial I $5p$ orbitals split to 3 groups of $a_{1g}$ (I−I bonding) and $e_g$ & $t_{1u}$ (I−I antibonding). The 12 tangential I $5p$ orbitals form 4 triply degenerated groups of $1t_{1u}$ & $t_{2g}$ (I−I bonding) and $t_{2u}$ & $t_{1g}$ (I−I antibonding). These groups are qualitatively arranged on the energy scale in Figure 6d.

By adding an Sn atom and 2 electrons (transferred from the two Cs atoms, which is ionized to $Cs^+$ in $Cs_2SnI_6$) into the $I_6$ octahedron, we calculated the electronic structure of a $[SnI_6]^{2-}$ octahedron cluster. The resulted DOSs are shown in Figure 6b and the derived energy levels are qualitatively illustrated in Figure 6d. The shallow Sn $5p$ orbitals slightly hybridize with the $I_6$ $t_{1u}$ orbitals, leading to Sn $5p$–$I_6$ $t_{1u}$ bonding states at −2.77 eV and Sn $5p$–$I_6$ $t_{1u}$ antibonding states at 5.45 eV. The deep Sn $5s$ orbitals strongly hybridize with the $I_6$ $a_{1g}$ orbitals (see the inset to Figure 6d), resulting in the Sn $5s$–$I_6$ $a_{1g}$ bonding states at −6.65 eV and the Sn $5s$–$I_6$ $a_{1g}$ antibonding states at 2.36 eV. These results are consistent with the −COHP analysis of Sn−I (the second panel of Figure 5a). The Sn $5s$ and $5p$ orbitals have little hybridization with the other 5 groups of I $5p$ orbitals (i.e. $t_{2g}$, $t_{2u}$, $e_g$, $t_{1g}$ and $2t_{1u}$). Note that the $2t_{1u}$ orbitals are fully occupied while the Sn $5s$–$I_6$ $a_{1g}$ antibonding orbital is unoccupied in the $[SnI_6]^{2-}$ cluster, and the $[SnI_6]^{2-}$ cluster forms a semiconductor-type electronic structure with the band gap of 2.36 eV.

Finally, we calculated a $[SnI_6]^{2-}$ sublattice model in which $[SnI_6]^{2-}$ are located at the corner and the face-center sites in the unit cell of $Cs_2SnI_6$ (i.e., the $Cs^+$ ions are removed from $Cs_2SnI_6$). The resulted DOSs (Figure 6c) are very similar to that of $Cs_2SnI_6$ (Figure 3b), indicating that $Cs^+$ cations have limited contribution to the electronic structure of $Cs_2SnI_6$ except for slightly pushing up the bands above $E_F$. The localized orbitals in the $[SnI_6]^{2-}$ cluster model expand into dispersed bands made of the $[SnI_6]^{2-}$ sublattice. The unoccupied Sn $5s$–$I_6$ $a_{1g}$ antibonding orbitals forms the CBM, and the occupied $I_6$ $5p$ ($t_{2g}$, $t_{2u}$, $e_g$, $t_{1g}$, and $2t_{1u}$) orbitals, which are antibonding between two I ions while non-bonding states to Sn, form the VBM.

## Conclusions

In conclusion, the real oxidation state of Sn in $Cs_2SnI_6$ is +2, which is the same as that in $CsSnI_3$ but different from the previously-expected value +4. DFT calculations clarified that the +2 oxidation state and the band gap are formed due to the intracluster hybridization in in the $[SnI_6]$ clusters and stabilized by the strong covalent nature of the Sn–I bonds. Further, this result indicate that the improved stability of $Cs_2SnI_6$ than $CsSnI_3$ is not attributed to the oxidation state of Sn because both $Cs_2SnI_6$ and $CsSnI_3$ have the formal oxidation state of 2+. Instead, the improved stability would be attributed to the shortened and stronger Sn–I length in $Cs_2SnI_6$ than that in $CsSnI_3$, which comes from the isolation of the $[SnI_6]^{2-}$ octahedra, similar to stable functional groups like $[SO_4]^{2-}$. The present result also shows the conventional ionic model is invalid for $p$-block metal-iodide based perovskites in which ionic bonds and covalent bonds coexist and compete, and provides a guiding principle to design new perovskite-based photovoltaic materials.

**Acknowledgements**

This work was conducted under Tokodai Institute for Element Strategy (TIES) by MEXT Elements Strategy Initiative to Form Core Research Center. Y.Z. thanks Prof. Nitin P. Padture and U.S. National Science Foundation (DMR-1305913) for the support.

**Tables and captions**

**Table 1.** Literature and calculated lattice parameters, bonding lengths and band gaps of $Cs_2SnI_6$.

|  | Exp. | LDA | PBE96 | HSE06 ($\alpha$=25%) | HSE06 ($\alpha$=34%) |
| --- | --- | --- | --- | --- | --- |
| $a$ (Å) | 11.65[23] | 11.26 | 12.03 | 11.86 | 11.82 |
| Cs–I (Å) | 4.12[23] | 3.98 | 4.25 | 4.19 | 4.18 |
| Sn–I (Å) | 2.85[23] | 2.83 | 2.91 | 2.89 | 2.88 |
| I–I (Å) | 4.04[23] | 4.00 | 4.11 | 4.09 | 4.07 |
| I–I' (Å) | 4.20[23] | 3.96 | 4.39 | 4.30 | 4.29 |
| $E_g$ (eV) | 1.26[13] | 0.13 | 0.25 | 0.93 | 1.26 |

**Table 2.** Literature bonding lengths, the ideal ionic ones estimated from the literature ion radii of 1.88 ($Cs^+$), 0.69 ($Sn^{4+}$), 1.02 ($Sn^{2+}$), and 2.20 Å ($I^-$) and the covalent ones from the proposed radius of 2.02 Å for $I^{-2/3}$ anion for $CsSnI_3$ and $Cs_2SnI_6$.

|  | $CsSnI_3$[10] | Ionic ($I^-$) | $Cs_2SnI_6$[23] | Covalent ($I^{-2/3}$) |
| --- | --- | --- | --- | --- |
| Cs–I (Å) | 4.39 | 4.22 | 4.12 | 3.90 |
| Sn–I (Å) | 3.10 | 3.22 ($Sn^{2+}$) | 2.85 | 3.04 ($Sn^{2+}$) |
|  |  | 2.89 ($Sn^{4+}$) |  | 2.71 ($Sn^{4+}$) |
| I–I (Å) | 4.39 | 4.40 | 4.04 | 4.04 |

**Figure captions**

**Figure 1.** (a) Unit cell and (b) 2×2×2 supercell of CsSnI$_3$. (c) Crystal structure of Cs$_2$SnI$_6$, which is obtained by removing a half of the Sn atoms at intervals (i.e. the edge centers and the body center of (b)). (d) Top view of (c). The bonding lengths from Ref. 23 are shown in (d).

**Figure 2.** Schematic electronic structure for (a) Cs$^+$Sn$^{2+}$I$^-_3$ (b) previously-expected Cs$^+_2$Sn$^{4+}$I$^-_6$, and (c) calculated Cs$^+_2$Sn$^{2+}$I$_6^{4-}$. The molecular orbital energy diagrams for Sn 5$s$–I 5$p$ bonds are also shown schematically on bottom-left in each figure.

**Figure 3.** (a) Band structure and (b) total and projected DOSs of Cs$_2$SnI$_6$ calculated with HSE06. The red and blue dashed lines mark the VBM and the CBM, respectively.

**Figure 4.** Bader oxidation states Z of Sn in Sn-based compounds.

**Figure 5.** (a) Calculated –COHPs for Sn−I, Cs−I, I−I, and I−I' bonds. The total DOS is shown in the top panel for comparison. The –ICOHP values for Sn−I, Cs−I, I−I, and I−I' are also shown in respective panels. (b and c) Valence electron density maps on the (b) (200) and (c) (400) planes.

**Figure 6.** Total and projected DOSs of (a) I$_6^0$ cluster, (b) [SnI$_6$]$^{2-}$ cluster, and (c) [SnI$_6$]$^{2-}$ sublattice models. The structures of the I$_6$ cluster, the SnI$_6$ cluster and the SnI$_6$ sublattice models are shown on top. (d) Qualitative interaction diagram for the I$_6^0$ cluster, the SnI$_6$ cluster, and the SnI$_6$ sublattice models, which is simplified from Figures (a) and (b) and the energy levels are qualitatively arranged on the energy scale. To easily understand the correspondence to the band structure of Cs$_2$SnI$_6$ (Figure 3), the energy level values are provided for the [SnI$_6$]$^{2-}$ cluster in (d), which are taken from the bonding states and the anti-bonding states in the band structure of Cs$_2$SnI$_6$. Inset of (d) shows a schematic illustration of I$_6$ a$_{1u}$ orbital.

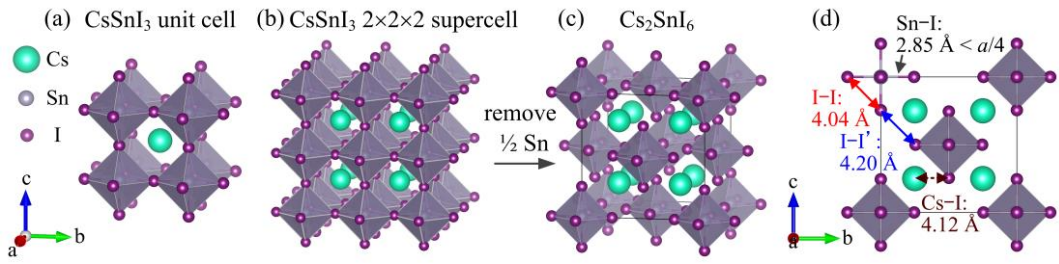

Figure 1. Xiao *et al.*

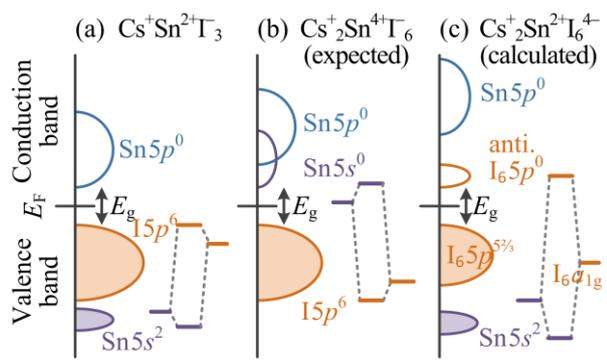

Figure 2. Xiao *et al.*

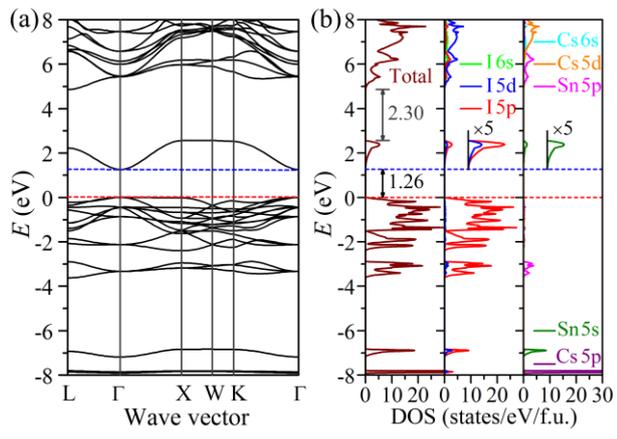

Figure 3. Xiao *et al.*

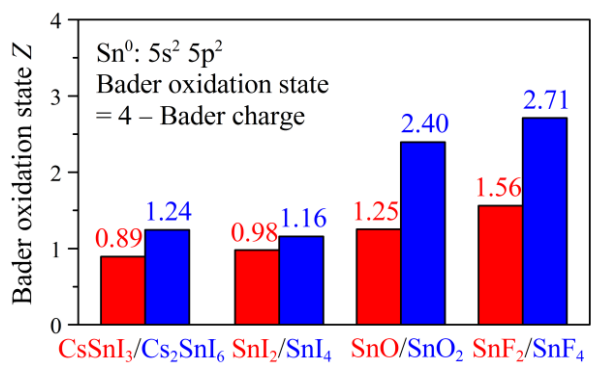

Figure 4. Xiao *et al.*

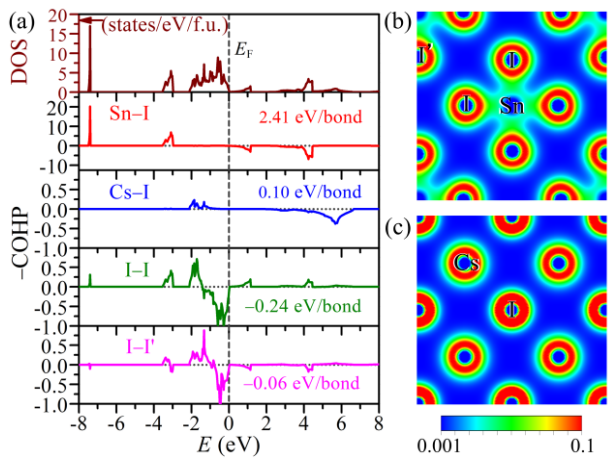

Figure 5. Xiao *et al.*

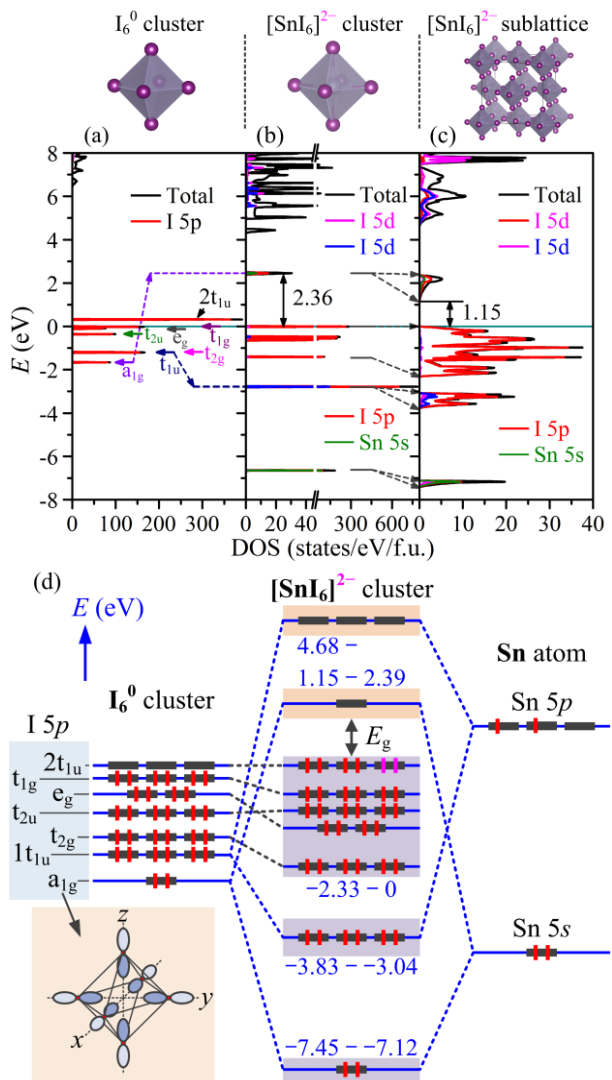

Figure 6. Xiao *et al.*

Graphical abstract

**Ligand-Hole in SnI$_6$ Unit and Origin of Band Gap in Photovoltaic Perovskite Variant Cs$_2$SnI$_6$**

Z. Xiao, H. Lei, X. Zhang, Y. Zhou, H. Hosono, and T. Kamiya

The real oxidation state of Sn in Cs$_2$SnI$_6$ is +2, rather than the expected "+4". The +2 oxidation state of Sn is realized by generating 2 ligand holes in the SnI$_6$ unit and stabilized by strong covalent Sn–I bonds.

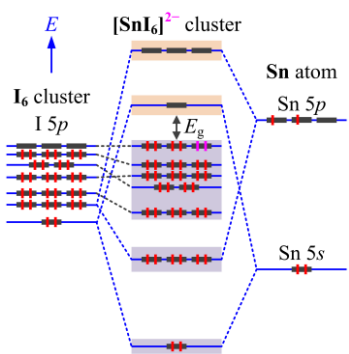